\documentclass[11pt, a4paper]{article}

\def\bSig\mathbf{\Sigma}

\usepackage[french, english]{babel} 
\usepackage{tikz} %
\usepackage{helvet} 
\usepackage{multirow} 
\usepackage{amsmath} 
\newcommand{\R}{\mathbb{R}} %
\usepackage{amsfonts, amssymb} 
\usepackage[figuresright]{rotating} 
\usepackage{subfig} 

\usepackage{url}

\usepackage{graphicx}
\usepackage[myheadings]{fullpage}
\usepackage{epsfig,epsf}
\usepackage{amssymb}
\usepackage[nolists]{endfloat}
\usepackage{setspace}
\usepackage{amsmath}

\usepackage{txfonts}

\usepackage{apacite}
\usepackage[sort&compress,authoryear]{natbib} 

\usepackage{prodint} 

\begin{document}

\title{Joint modelling of longitudinal and multi-state processes:\\ application to clinical progressions in prostate cancer}

\author{Lo\"ic Ferrer$^{1,*}$, Virginie Rondeau$^{1}$, James J. Dignam$^{2}$, Tom Pickles$^{3}$, \\H\'el\`ene Jacqmin-Gadda$^{1}$ and C\'ecile Proust-Lima$^{1}$}

\date{June 2015}

\maketitle

\vspace{0.3cm}

\noindent 
$^{1}$ INSERM, U897, ISPED, Universit\'e de Bordeaux, F-33000 Bordeaux, France \\
$^{2}$ Department of Public Health Sciences, University of Chicago, and NRG Oncology, U.S.A. \\
$^{3}$ Department of Radiation Oncology, University of British Columbia, Canada \\
$^{*}$ E-mail: loic.ferrer@inserm.fr

\label{firstpage}

\vspace{0.9cm}
\noindent {\bf {Abstract:}}
Joint modelling of longitudinal and survival data is increasingly used in clinical trials on cancer. In prostate cancer for example, these models permit to account for the link between longitudinal measures of prostate-specific antigen (PSA) and the time of clinical recurrence when studying the risk of relapse.
In practice, multiple types of relapse may occur successively. Distinguishing these transitions between health states would allow to evaluate, for example, how PSA trajectory and classical covariates impact the risk of dying after a distant recurrence post-radiotherapy, or to predict the risk of one specific type of clinical recurrence post-radiotherapy, from the PSA history.
In this context, we present a joint model for a longitudinal process and a multi-state process which is divided into two sub-models: a linear mixed sub-model for longitudinal data, and a multi-state sub-model with proportional hazards for transition times, both linked by shared random effects.
Parameters of this joint multi-state model are estimated within the maximum likelihood framework using an EM algorithm coupled to a quasi-Newton algorithm in case of slow convergence.
It is implemented under {R}, by combining and extending the \texttt{mstate} and \texttt{JM} packages.
The estimation program is validated by simulations and applied on pooled data from two cohorts of men with localized prostate cancer and treated by radiotherapy.
Thanks to the classical covariates available at baseline and the PSA measurements collected repeatedly during the follow-up, we are able to assess the biomarker's trajectory, define the risks of transitions between health states, and quantify the impact of the PSA dynamics on each transition intensity.\\


\pagebreak

\doublespace

\section{Introduction}
\label{s:intro}

In longitudinal health studies, marker data are usually collected at repeated measurement times until the occurrence of an event such as disease relapse or death, with the perspective to study the link between these two correlated processes or use the information brought by the marker's dynamics to explain or predict the time to event.
In such analyses, the repeated measurements of the marker should not be considered as a standard time-dependent covariate in a survival model \citep{andersen1982cox, fisher1999time} because the marker is an internal outcome measured with error and at discrete times whereas the Cox model assumes to know the exact values of the explanatory variables for all individuals at risk at each event time. 
To counteract these weaknesses, the two processes can be modelled jointly \citep{faucett1996simultaneously, wulfsohn1997joint}.
The principle is to define two sub-models (one mixed sub-model for the longitudinal data and one survival sub-model for the time-to-event data) and link them using a common latent structure.
The shared random effect models, notably developed by \citet{tsiatis2004joint}, are the most popular joint models. They usually assume that a function of the random effects from the linear mixed model is included as covariates in the survival model. Thus, longitudinal and survival processes share the same random effects, and the effect of the dynamics of the marker on the risk of event can be deduced.

In prostate cancer, the joint modelling method is very useful. The prostate-specific antigen (PSA), which is a protein secreted by the prostate, is found to be over-expressed in the presence of prostate cancer. This blood-based longitudinal tumour marker is commonly used by clinicians to monitor patients with localized prostate cancer following treatment (radiation therapy or surgery) in order to detect subclinical presence of disease. \citet{proust2008determinants} , \citet{taylor2005individualized} and \citet{yu2008individual} showed, through various types of joint models, that the dynamics of this biomarker, along with the pre-treatment PSA level and other factors measuring the aggressiveness of cancer cells and the extent of the tumour, were risk factors for progression and permitted one to dynamically predict (i.e. using PSA to adapt prediction over time) the risk of clinical relapse.

In practice, a patient may experience a succession of clinical progression events with for example a local recurrence, followed by a distant metastatic recurrence and then death. So, instead of the occurrence of a single clinical event, the progression of prostate cancer should be defined as a multi-state process with a focus on the transitions between clinical states and the impact of the biomarker dynamics on it. This is essential to understand and predict accurately the course of the disease, and it is of particular relevance for the clinicians that need to distinguish the different types of events in order to properly adapt the treatment. 

Some authors already extended the classical framework of joint modelling to multiple time-to-event data.
\citet{chi2006joint} proposed a joint model for multivariate longitudinal data and multivariate survival data. 
\citet{liu2009joint} and \citet{kim2012joint} looked into the simultaneous study of three correlated outcomes: longitudinal data, times of recurrent events and time of terminal event.
\citet{elashoff2008joint} and \citet{rizopoulos2012joint} extended the joint model to competing risks survival data, which allows to characterize the cause of survival event. 
\citet{dantan2011joint} presented a joint model with latent state for longitudinal data and time-to-illness and time-to-death data.
Recently, \citet{andrinopoulou2014joint} studied simultaneously two longitudinal markers and competing events.
However, the joint study of longitudinal and multi-state data has never been proposed and implemented. 
Thus, we introduce a joint model with shared random effects for repeated measurements of a longitudinal marker and times of transitions between multiple states. It consists in a linear mixed model and a multi-state model with transition-specific proportional intensities, both linked by shared random effects.

The computational aspect is the main obstacle in the development of joint models with shared random effects. As explained by \citet{lawrence2014joint}, the {R} package \texttt{JM}, developed by \citet{rizopoulos2010jm}, has enabled many advances in the use of joint modelling, particularly through efficient numerical integrations.
On the other hand, the {R} package \texttt{mstate}, developed by \citet{dewreede2010mstate}, provides estimation of multi-state models. 
In the present work, we combine and adapt these two packages in order to estimate joint multi-state models.
Thus, the implementation is easy and effective.
Through the adaptation of the \texttt{jointModel()} function of the \texttt{JM} package, our approach uses the maximum likelihood approach, which is performed using the EM algorithm coupled to a quasi-Newton algorithm in case of slow convergence. 
The software advantage is that it keeps the features, syntax and outputs of \texttt{JM}. 

The paper is organized as follows. Section~\ref{s:joint} presents the joint model for longitudinal and multi-state processes. The estimation and implementation procedures are detailed in Section~\ref{s:estimation}. The joint multi-state model has been validated by simulations, as detailed in Section~\ref{s:simulation}. We apply our model on a detailed example of two cohorts of men with prostate cancer in Section~\ref{s:application}. A brief discussion is given in Section~\ref{s:discussion}.

\section{Joint multi-state model}
\label{s:joint}

\subsection{Notations}
\label{s:s:notations}

For each individual $i$, a longitudinal process and a multi-state process are observed.
Let $\lbrace E_i(t), t \geq 0 \rbrace$ be the multi-state process where $E_i(t)$ denotes the occupied state by subject $i$ at time $t$ and takes values in the finite state space $S=\lbrace 0, 1, \ldots, M \rbrace$.
It is assumed that the multi-state process is continuous and observed between the left truncation time $T_{i0}$ and the right censoring time $C_i$, so that the observed process is $E_i = \lbrace E_i(t), T_{i0} \leq t \leq C_i \rbrace$.
We further consider that $E_i$ is a non-homogeneous Markov process. The Markov property ensures that the future of the process depends only on the present state and not the past state, i.e. $\Pr \left( E_i(t + u) = k | E_i(t) = h, \{E_i(s), s<t \} \right) = \Pr \left( E_i(t + u) = k | E_i(t) = h \right), \forall h,k \in S, \forall u \geq 0$ (\citet{dewreede2010mstate}), and the non-homogeneous property guarantees that the time since $T_{i0}$ impacts the future evolution of the process.
Let us consider $T_i = \left( T_{i1}, T_{i2}, \ldots, T_{im_i} \right)^\top$ the vector of the $m_i \geq 1$ observed times for the individual $i$, with $T_{ir} < T_{i(r+1)}, \forall r \in \lbrace 0, \ldots, m_i-1\rbrace$, and where $^\top$ denotes the transpose vector.
If the last observed state for subject $i$ ($E_i(T_{im_i})$) is absorbing, that is it is impossible to leave it once entered (typically death), we observe $m_i$ direct transitions. Otherwise, $T_{im_i}$ equals $C_i$ the right censoring time and we observe $m_i - 1$ direct transitions.
We define by $\delta_i = \left(\delta_{i1}, \ldots, \delta_{im_i} \right)^\top$ the vector of observed transition indicators, with $\delta_{i(r+1)}$ equals $1$ if a transition is observed at time $T_{i(r+1)}$ (i.e. $E_i(T_{ir}) \neq E_i(T_{i(r+1)})$) and $0$ otherwise, $\forall r \in \lbrace 0, \ldots, m_i-1\rbrace$.
For each patient $i$, we also observe $Y_i = \left( Y_{i1}, \ldots, Y_{in_i} \right)^\top$ the vector of $n_i$ measures of the marker collected at times $t_{i1}, \ldots, t_{in_i}$, with $t_{in_i} \leq T_{im_i}$.

\subsection{Joint multi-state model formulation}
\label{s:s:model}

The joint multi-state model is decomposed into two sub-models : a linear mixed sub-model for the longitudinal data (repeated measurements of the biomarker) and a multi-state model with proportional hazards for the event history data (transition and censoring times), both linked by shared random effects.

\subsubsection{Longitudinal sub-model}
\label{s:s:s:longit}

To model the trajectory of the longitudinal marker, we use a linear mixed model. Under Gaussian assumptions, we assume that $Y_{ij}$ the observed measure of the marker at time point $t_{ij}$ is a noisy measure of the true measure $Y_{i}^{*}(t_{ij})$. This non-observed measure $Y_{i}^{*}(t_{ij})$ is explained according to time and covariates, at the population level, with fixed effects $\beta$ and at the individual level, with random effects $b_i$ that take into account the correlation between repeated measures of the same individual:
\begin{eqnarray}
Y_{ij} & = &Y_{i}^{*}(t_{ij}) + \epsilon_{ij} \nonumber \\
	   & = &X^{L}_{i}(t_{ij})^{\top}\beta + Z_{i}(t_{ij})^{\top}b_i + \epsilon_{ij} ,
\label{e:mixed}
\end{eqnarray}
with $X^{L}_{i}(t_{ij})$ and $Z_{i}(t_{ij})$ the vectors of possibly time-dependent covariates associated respectively with the $p$-vector of fixed effects $\beta$ and the $q$-vector of random effects $b_i$, $b_i \sim \mathcal{N}(0, D)$. Note that $\epsilon_{i} = \left( \epsilon_{i1}, \ldots, \epsilon_{in_i} \right)^\top \sim \mathcal{N}(0, \sigma^2 I_{n_i})$ where $I$ is the identity matrix; $\epsilon_i$ and $b_i$ are independent.

\subsubsection{Multi-state sub-model}
\label{s:s:s:multistate}

To model the transition times, we use a Markov multi-state model with proportional hazards that takes into account the marker's dynamics through the shared random effects $b_i$.
Thus, for the transition from state $h \in S$ to state $k \in S$, the transition intensity at time $t$ takes the form:
\begin{eqnarray}
\label{e:mstate}
\lambda^{i}_{hk}(t | b_i) & = &\lim_{\mathrm dt \rightarrow 0} \dfrac{\Pr(E_i(t + \mathrm dt)=k | E_i(t)=h; b_i)}{\mathrm dt} \nonumber \\
						& = &\lambda_{hk,0}(t)\exp({ X^{S~\top}_{hk,i} \gamma_{hk} + W_{hk,i}(b_i,t)^{\top} \eta_{hk} }) ,
\end{eqnarray}
with $\lambda_{hk,0}(.)$ the parametric baseline intensity (Weibull, piecewise constant or B-splines for example) and $X^{S}_{hk,i}$ the vector of prognostic factors associated with the $r$-vector of coefficients $\gamma_{hk}$.
The multivariate function $W_{hk,i}(b_i, t)$ defines the dependence structure between the longitudinal and multi-state processes. We can choose $W_{hk,i}(b_i, t) = Y^{*}_{i} (t)$ (the true current level of the marker), or $W_{hk,i}(b_i,t) = \partial Y_{i}^{*}(t)/\partial t$ (the true current slope), $W_{hk,i}(b_i, t)=\left( Y_{i}^{*}(t),\partial Y_{i}^{*}(t) / \partial t \right)^\top$ (both), or any other function of the random effects in the context under study. Thus, the $s$-vector of coefficients $\eta_{hk}$ quantifies the impact of the longitudinal marker's dynamics on the transition intensity between states $h$ and $k$.

\section{Estimation}
\label{s:estimation}

\subsection{Likelihood}
\label{s:s:likelihood}

The parameters of this joint model are estimated in the maximum likelihood framework.
Since the longitudinal and multi-state processes are independent conditionally on the random effects, the complete observed likelihood is obtained through the product of the individual contributions to the likelihood for the $N$ individuals as:
\begin{eqnarray}
\label{e:likelihood_joint_indiv}
L(\theta) = \prod_{i=1}^{N} \int_{\R^{q}} f_{Y}(Y_i|b_i;\theta) f_{E}(E_i|b_i;\theta) f_{b}(b_i;\theta) \, \mathrm db_i ,
\end{eqnarray}
where $\theta$ is the vector of all the parameters contained in~(\ref{e:mixed}) and~(\ref{e:mstate}), and $f(.)$ is a probability density function.

In the longitudinal sub-part, described by the linear mixed model~(\ref{e:mixed}), the conditional longitudinal outcomes are such that:
\begin{eqnarray}
\label{e:likelihood_longit}
f_{Y}(Y_i | b_i;\theta) = \dfrac{1}{\left( 2 \pi \sigma^2 \right)^{n_i/2}} \exp\left( -\frac{\|Y_i - X^{L~\top}_{i}\beta - Z_{i}^{\top} b_i\|^2}{2\sigma^2}\right) ,
\end{eqnarray}
where $||x||$ denotes the Euclidean norm of vector $x$, $X^{L}_{i}$ is the matrix of covariates with row vectors $X^{L}_{i}(t_{ij})^T$, $j=1,\ldots,n_i,$ and likewise $Z_i = \lbrace Z_{i}(t_{ij}) \rbrace$.

For the multi-state side, let $P^{i}_{hk}(s,t)$ be the transition probability from state $h$ to state $k$ between the times $s$ and $t$ for individual $i$, i.e. $P^{i}_{hk}(s,t) = \mathrm{Pr}(E_i(t)=k|E_i(s)=h)$.
For each $r \in \{0, \ldots, m_i - 1\}$, the continuity and Markov assumptions imply that the individual $i$ remains in state $E_i(T_{ir})$ between the times $T_{ir}$ and $T_{i(r+1)}$ with probability $P^i_{E_i(T_{ir}), E_i(T_{ir})}(T_{ir},T_{i(r+1)} | b_i)$, and if $T_{i(r+1)}$ is an observed transition time, he transits to state $E_i(T_{i(r+1)})$ with intensity $\lambda^i_{E_i(T_{ir}), E_i(T_{i(r+1)})} (T_{i(r+1)} | b_i)$.
By conditioning on $E_i(T_{i0})$, this translates in the individual contribution to the likelihood:
\begin{alignat}{4}
\label{e:likelihood_mstate}
f_{E}(E_i | b_i;\theta)& = &\prod_{r=0}^{m_i - 1} \bigg[ &P^i_{E_i(T_{ir}), E_i(T_{ir})}(T_{ir},T_{i(r+1)} | b_i) \, {\lambda^i_{E_i(T_{ir}), E_i(T_{i(r+1)})} (T_{i(r+1)} | b_i)}^{\delta_{i(r+1)}} \bigg] \nonumber \\
	& = &\prod_{r=0}^{m_i - 1} \bigg[ &\exp\left( \int_{T_{ir}}^{T_{i(r+1)}} \lambda^i_{E_i(T_{ir}), E_i(T_{ir})}(u|b_i) \, \mathrm du \right) {\lambda^i_{E_i(T_{ir}), E_i(T_{i(r+1)})} (T_{i(r+1)} | b_i)}^{\delta_{i(r+1)}} \bigg]
\end{alignat}
with $\lambda^{i}_{hh}(t) = - \sum_{k, k \neq h} \lambda^{i}_{hk}(t)$.
The possible delayed entry is here ignored by conditioning on $E_i(T_{i0})$.

Linking the longitudinal and multi-state sub-parts, the random effects $b_i$ follow a multivariate Gaussian distribution such that:
\begin{eqnarray}
\label{e:likelihood_random}
f_{b}(b_i;\theta) = \dfrac{1}{ {\left( 2 \pi \right)}^{q/2} {\det(D)}^{1/2} } \exp\left( -\frac{b_{i}^{\top} D^{-1} b_i}{2}\right) .
\end{eqnarray}

\subsection{Implementation}
\label{s:s:implementation}

The joint multi-state model has been implemented under R, via the combination of two well-known packages: \texttt{mstate} for the multi-state models and \texttt{JM} for the joint models with shared random effects.
To fit semi-parametric Markov multi-state models, \texttt{mstate} consists first of preparing the database for multi-state analysis, more specifically by defining each patient's history as a series of rows, one for each transition at risk for each individual (in contrast with only one data record (row) per individual in a classical survival analysis).
By stratifying on the transition type, the standard \texttt{coxph()} function of the R package \texttt{survival} can be used to fit transition-specific Cox models.
With standard longitudinal and time-to-event data, the \texttt{JM} package initialises the values of parameters from the function \texttt{lme()} (\texttt{nlme} package) for the longitudinal sub-part and \texttt{coxph()} (\texttt{survival} package) for the survival sub-part. Then, the function \texttt{jointModel()} carries out the estimation procedure.

So by replacing the standard call to \texttt{coxph()} by the call to \texttt{coxph()} on prepared data by \texttt{mstate}, an extended \texttt{jointModel()} function, called \texttt{JMstateModel()}, carries out the estimation procedure. The implementation procedure is thus distinguished in four steps:
\begin{itemize}
\item \texttt{lme()} function to initialise the parameters in the longitudinal sub-model;
\item \texttt{mstate} package to adapt the data to the multi-state framework;
\item \texttt{coxph()} function, on the prepared data, to initialise the parameters in the multi-state sub-model;
\item \texttt{JMstateModel()} function to estimate all the parameters of the joint multi-state model.
\end{itemize}
A detailed example is given in Appendix C, and full detailed examples are available on \path{https://github.com/LoicFerrer/JMstateModel/}.

\subsection{Algorithm}
\label{s:s:algorithm}

The \texttt{JMstateModel()} function computes and maximises the likelihood adapted to the multi-state data using integration and optimisation algorithms available in the \texttt{JM} package.
Thus, the procedure combines an EM algorithm coupled to a quasi-Newton algorithm if the convergence is not achieved.
Furthermore, the integral with respect to time in~(\ref{e:likelihood_mstate}) and the integral with respect to the random effects in~(\ref{e:likelihood_joint_indiv}) do not have an analytical solution.
These integrals are approached by numerical integration. 
The integrals over time are approximated using Gauss-Kronrod quadratures, and the integrals over the random effects using pseudo-adaptative Gauss-Hermite quadratures.
Inference is provided by asymptotic properties for maximum likelihood estimators; the variance-covariance matrix of the parameter estimates is based on the inverse of the Hessian matrix.
More details on the optimisation procedure, the EM algorithm and the numerical integrations are available in \cite{rizopoulos2012joint}.

\section{Simulation study}
\label{s:simulation}

The simulation study aimed to validate the estimation program described previously.

\subsection{Data generation}
\label{s:s:data_generation}

For each subject $i=1,\ldots,1500$ of the 500 replicates, the longitudinal and multi-state data were generated according to the joint multi-state model defined as:
\begin{eqnarray}
\label{e:model_simu}
\left\{
\begin{array}{rll}
Y_{ij} & = &Y^{*}_{i}(t_{ij}) + \epsilon_{ij} \\
	& = &(\beta_0 + \beta_{0, X} X_i + b_{i0}) + (\beta_1 + \beta_{1, X} X_i + b_{i1}) \times t_{ij} + \epsilon_{ij} , \\
\lambda^{i}_{hk}(t | b_i) & = &\lambda_{hk,0}(t)\exp \left( \gamma_{hk} X_i + \eta_{hk, \texttt{level}} Y_{i}^{*}(t) + \eta_{hk,\texttt{slope}} \partial Y_{i}^{*}(t) / \partial t \right) ,
\end{array}
\right.
\end{eqnarray}
where the multi-state process that included three states ($h, k \in \{0,1,2\}$) and three transitions is described in Figure~\ref{f:simul_mstate}.
\begin{figure}
\begin{center}
\begin{tikzpicture}[scale=0.85]
  \tikzstyle{noeud}=[minimum width=2cm,minimum height=1.33cm,line  width=1pt, rectangle,draw] 

  \node[noeud] (st0) at (0,0) {\large State 0};
   \draw (st0.base) node[below=10pt, right=17pt]{\footnotesize 0};

  \node[noeud] (st1) at (6,0) {\large State 1};
   \draw (st1.base) node[below=10pt, right=17pt]{\footnotesize 1};
  \node[noeud] (st2) at (3,-4.5) {\large State 2};
   \draw (st2.base) node[below=10pt, right=17pt]{\footnotesize 2};
  \draw[->,>=latex,line width=1pt] (st0) -- (st1) node[midway,above]{\small $\lambda_{01}(t)$ } ;
  \draw[->,>=latex,line width=1pt] (st0) -- (st2);
  \draw[->,>=latex,line width=1pt] (st1) -- (st2);

 \draw (0.7,-2.25) node {\small $\lambda_{02}(t)$};
 \draw (5.3,-2.25) node {\small $\lambda_{12}(t)$};
 
  \draw (-1,-7) node[right, text width=5cm] {$\boldsymbol{\overline{\Upsilon}_{\textrm{sim}}} = \begin{pmatrix} 692 & 500 & 308 \\ 0 & 213 & 287 \\ 0 & 0 & 595 \end{pmatrix}$} ;
\end{tikzpicture}
\caption{Simulated multi-state process. Arrows indicate the directions of the possible transitions. $\lambda_{hk}(t)$ characterizes the intensity of transition between states $h$ and $k$ at time $t$. The matrix $\boldsymbol{\overline{\Upsilon}_{\textrm{sim}}}$ has size $(3, 3)$ and is composed of elements $\overline{\Upsilon}_{\textrm{sim}, (h+1)(k+1)}$, where $\overline{\Upsilon}_{\textrm{sim}, (h+1)(k+1)}$ is the average number of observed direct transitions $h \rightarrow k$ over the 500 replicates. The diagonal elements $\overline{\Upsilon}_{\textrm{sim}, (h+1)(h+1)}$ denote the average number of patients who were censored in state $h$. Note that the sum of elements of a row $(h+1)$ of $\overline{\Upsilon}_{\textrm{sim}}$ corresponds to the average number of patients who experienced the state $h$.}
\label{f:simul_mstate}
\end{center}
\end{figure}

First, $X_i$ and $b_i$ were generated according to normal distributions with respectively mean $2.04$ and variance $0.5$, and mean vector $\begin{pmatrix} 0 \\ 0 \end{pmatrix}$ and variance-covariance matrix $\begin{pmatrix} 0.35 & - 0.04 \\ -0.04 & 0.06 \end{pmatrix}$.
The times of measurements were generated such as $t_{ij} = 0, 0.33, 0.67, \ldots, 16.33$, and $\epsilon_{ij}$ was generated from a standard normal distribution.
The log baseline intensities were linear combinations of cubic B-splines with the same knot vector $(0.004, 4.120, 7.455, 10.908, 18.201)^\top$ for the three transitions, and the vectors of spline coefficients $(-9.200, -3.500, -5.000$, $-3.900, -3.500, -2.500, -2.000)^T$ for the transition $0 \rightarrow 1$, $(-9.860, -4.472, -5.128, -3.486$, $-2.457, -0.989, -0.715)^\top$ for the transition $0 \rightarrow 2$, and $(-2.527, -2.170, -2.492, -2.156$, $-1.228, -0.955, -0.161)^\top$ for the transition $1 \rightarrow 2$.
Parameters values and knot locations were chosen according to the application data described in Section~\ref{s:application}.

The procedure to generate the vector of observed times $T_i = \left(T_{i1}, \ldots, T_{im_i} \right)^\top$ was inspired by \citet{beyersmann2011competing} and \cite{crowther2013simulating}.
For each individual $i$, the censoring time $C_i$ was generated from an uniform distribution on $\left[1, 25\right]$, and the vector of true transition times $T_{i}^{*} = \left( T_{i,01}^{*}, T_{i,02}^{*}, T_{i,12}^{*} \right)^\top$ was generated according to the following procedure: (1) three random numbers $u_{i,01}$, $u_{i,02}$ and $u_{i,12}$ were generated from three independent standard uniform distributions;
(2) $T_{i,01}^{*}$ and $T_{i,02}^{*}$ were generated by solving $\int_{0}^{T_{i,0k}^{*}} \lambda^{i}_{0k}(\nu_{0k}|b_i) \, \mathrm d\nu_{0k} + \log(u_{i,0k}) = 0 ,$ for $k=1,2 ,$ through the Brent's univariate root-finding method \citep{brent1973algorithms};
(3) then, the true transition time $T_{i,12}^{*}$ was generated via $\int_{T_{i,01}^{*}}^{T_{i,12}^{*}} \lambda^{i}_{12}(\nu_{12}|b_i) \, \mathrm d\nu_{12} + \log(u_{i,12}) = 0$.
Finally, by comparing $T_{i}^{*}$ and $C_i$, the vector $T_i$, which characterizes the multi-state process, was deduced.

The longitudinal measurements, generated from the linear mixed sub-model, were truncated at $T_{i1}$ the first observed time of the multi-state process.

\subsection{Estimated model}
\label{s:s:estimated_model}

The estimation model was the model defined in~(\ref{e:model_simu}) with $b_i \sim \mathcal{N}\left(\begin{pmatrix}
0 \\ 0 \end{pmatrix}, \begin{pmatrix} D_{11} & D_{12} \\ D_{12} & D_{22} \end{pmatrix} \right)$ and $\epsilon_{ij} \sim \mathcal{N}(0, \sigma^2)$. The log baseline intensities were approximated by a linear combination of cubic-splines with three internal knots placed at the quantiles of the observed times.

\subsection{Simulation results}
\label{s:s:simulation_results}

The simulations results were obtained through 500 replicates of 1500 individuals.
Each joint multi-state model was estimated using 3, 9 and 15 pseudo-adaptative Gauss-Hermite quadrature points. The simulation results are presented in Table~\ref{t:simulation_results}.

\begin{sidewaystable}[f]
\centering
\begin{small}
\caption[Simulation results according to 3, 9 and 15 pseudo-adaptative Gauss-Hermite quadrature points. For each scenario, the statistics are (from left to right): mean,  mean standard error, standard deviation, relative bias, root-mean-square-error and coverage rate.]{Simulation results according to 3, 9 and 15 pseudo-adaptative Gauss-Hermite quadrature points. For each scenario, the statistics are (from left to right): mean,  mean standard error, standard deviation, relative bias (in percentage) and coverage rate (in percentage).}
\label{t:simulation_results}
\begin{tabular}{lrrrrrrrrrrrrrrrrrrr}
  \hline
 & & \multicolumn{5}{c}{3 Gauss-Hermite quadrature points} & \multicolumn{5}{c}{9 Gauss-Hermite quadrature points} & \multicolumn{5}{c}{15 Gauss-Hermite quadrature points} \\
 & & \multicolumn{5}{c}{\hrulefill} & \multicolumn{5}{c}{\hrulefill} & \multicolumn{5}{c}{\hrulefill} \\
 & True & \multirow{2}{*}{Mean} & \multirow{2}{*}{$\overline{\textrm{StdErr}}$} & \multirow{2}{*}{StdDev} & Rel. & Cov. & \multirow{2}{*}{Mean} & \multirow{2}{*}{$\overline{\textrm{StdErr}}$} & \multirow{2}{*}{StdDev} & Rel. & Cov. & \multirow{2}{*}{Mean} & \multirow{2}{*}{$\overline{\textrm{StdErr}}$} & \multirow{2}{*}{StdDev} & Rel. & Cov. \\
  & value & & & & bias & rate & & & & bias & rate & & & & bias & rate \\
  \hline
\multicolumn{2}{l}{\hspace{-0.35cm} \textit{Longitudinal process}} &&&&&&&&&&&&&&& \\
$\beta_0$ & $-$0.793 & $-$0.797 & 0.049 & 0.050 & 0.4 & 95.8 & $-$0.796 & 0.049 & 0.050 & 0.3 & 95.8 & $-$0.796 & 0.049 & 0.050 & 0.3 & 95.8 \\
  $\beta_{0, X}$ & 0.543 & 0.545 & 0.023 & 0.023 & 0.3 & 94.8 & 0.544 & 0.023 & 0.023 & 0.2 & 95.0 & 0.544 & 0.023 & 0.023 & 0.2 & 95.2 \\
  $\beta_1$ & $-$0.096 & $-$0.093 & 0.015 & 0.021 & $-$3.4 & 85.6 & $-$0.096 & 0.021 & 0.021 & $-$0.3 & 94.8 & $-$0.096 & 0.021 & 0.021 & $-$0.3 & 94.8 \\
  $\beta_{1, X}$ & 0.027 & 0.024 & 0.006 & 0.010 & $-$10.4 & 78.4 & 0.026 & 0.010 & 0.010 & $-$1.1 & 94.0 & 0.026 & 0.010 & 0.010 & $-$1.1 & 94.2 \\
  $\log(\sigma)$ & $-$0.737 & $-$0.737 & 0.004 & 0.004 & 0.0 & 94.2 & $-$0.737 & 0.004 & 0.004 & 0.0 & 94.2 & $-$0.737 & 0.004 & 0.004 & 0.0 & 94.0 \\
\multicolumn{2}{l}{\hspace{-0.35cm} \textit{Multi-state process}} &&&&&&&&&&&&&&& \\
  $\gamma_{01, X}$ & 0.281 & 0.289 & 0.077 & 0.078 & 2.6 & 95.0 & 0.292 & 0.077 & 0.078 & 3.7 & 95.0 & 0.292 & 0.077 & 0.078 & 3.7 & 95.2 \\
  $\gamma_{02, X}$ & 0.023 & 0.022 & 0.088 & 0.090 & $-$4.2 & 94.8 & 0.022 & 0.088 & 0.089 & $-$5.4 & 95.6 & 0.024 & 0.088 & 0.089 & 2.7 & 94.6 \\
  $\gamma_{12, X}$ & $-$0.169 & $-$0.183 & 0.095 & 0.096 & 8.0 & 94.8 & $-$0.175 & 0.095 & 0.096 & 3.6 & 94.4 & $-$0.175 & 0.095 & 0.095 & 3.6 & 95.6 \\
  $\eta_{01, \texttt{level}}$ & 0.925 & 0.918 & 0.073 & 0.074 & $-$0.7 & 93.6 & 0.912 & 0.073 & 0.074 & $-$1.3 & 93.6 & 0.912 & 0.073 & 0.073 & $-$1.4 & 93.8 \\
  $\eta_{02, \texttt{level}}$ & 0.297 & 0.301 & 0.064 & 0.063 & 1.5 & 95.0 & 0.299 & 0.064 & 0.065 & 0.6 & 94.8 & 0.298 & 0.064 & 0.064 & 0.4 & 95.0 \\
  $\eta_{12, \texttt{level}}$ & 0.071 & 0.077 & 0.074 & 0.074 & 8.7 & 94.4 & 0.071 & 0.074 & 0.074 & 0.1 & 94.4 & 0.071 & 0.074 & 0.074 & $-$0.5 & 94.2 \\
  $\eta_{01, \texttt{slope}}$ & 1.344 & 1.456 & 0.436 & 0.427 & 8.3 & 94.6 & 1.502 & 0.439 & 0.437 & 11.7 & 92.6 & 1.504 & 0.439 & 0.436 & 11.9 & 93.4 \\
  $\eta_{02, \texttt{slope}}$ & $-$1.096 & $-$1.123 & 0.640 & 0.623 & 2.5 & 96.0 & $-$1.093 & 0.641 & 0.642 & $-$0.3 & 94.0 & $-$1.085 & 0.642 & 0.634 & $-$1.0 & 95.0 \\
  $\eta_{12, \texttt{slope}}$ & 0.009 & $-$0.061 & 0.788 & 0.789 & $-$777.3 & 94.8 & 0.008 & 0.790 & 0.801 & $-$7.3 & 94.6 & 0.016 & 0.790 & 0.795 & 72.3 & 95.0 \\
\multicolumn{2}{l}{\hspace{-0.35cm} \textit{Random effects}} &&&&&&&&&&&&&&& \\
  $D_{11}$ & 0.349 & 0.348 & 0.014 & 0.014 & $-$0.3 & 96.0 & 0.348 & 0.014 & 0.014 & $-$0.3 & 96.0 & 0.348 & 0.014 & 0.014 & $-$0.3 & 96.2 \\
  $D_{12}$ & $-$0.041 & $-$0.042 & 0.005 & 0.004 & 0.5 & 95.8 & $-$0.042 & 0.005 & 0.004 & 0.6 & 95.8 & $-$0.042 & 0.005 & 0.004 & 0.6 & 95.8 \\
  $D_{22}$ & 0.062 & 0.062 & 0.003 & 0.003 & $-$0.1 & 91.2 & 0.062 & 0.003 & 0.003 & $-$0.0 & 91.8 & 0.062 & 0.003 & 0.003 & 0.0 & 92.0 \\
   \hline \\
\end{tabular}
\end{small}
\end{sidewaystable}

These results were very satisfying with unbiased estimates and correct 95\% coverage rates. They showed however the need to use a certain number of Gauss-Hermite quadrature points to approximate the integral over the random effects. Indeed, the use of 3 pseudo-adaptative Gauss-Hermite quadrature points induced a bias in the estimation of the parameters associated with time effect in the longitudinal sub-part. This error was fully corrected with 9 points. Overall, these results confirmed the good performances of the implemented procedure.
Complementary simulation results based on 500 and 1000 subjects are provided in Appendix A.

\section{Application}
\label{s:application}

We analysed data from patients with localized prostate cancer and treated by external beam radiotherapy. The analysis aimed to explore the link between PSA dynamics and transition intensities between clinical states, as well as to describe PSA repeated measurements and times of transitions between health states.

\subsection{Data description}
\label{s:s:data}

Our study focuses on 1474 men with a clinically localized prostate cancer and treated by external beam radiotherapy (EBRT): 629 patients come from the multi-center clinical trial RTOG 9406 (Radiation Therapy Oncology Group, USA) in which data collection has been conducted from 1994 to 2013, and 845 patients come from the cohort of the British Columbia Cancer Agency (BCCA) in Vancouver, Canada, and have been examined between 1994 and 2012 (Table~\ref{t:description_cohorts}).
During his follow-up, a patient can possibly go through several states defined as local recurrence, distant recurrence, initiation of hormonal therapy (HT) and death, due or not to prostate cancer.
The initiation of salvage hormonal therapy, which is an additional treatment prompted by physician observed signs in PSA or clinical signs, is designed to prevent growth of potentially present subclinical cancer. This intervention is not planned at diagnosis or initiated by any precise rule, but is rather based on a mutual agreement between the clinician and his patient. Thus, it is treated as a disease state transition representing failure of the initial treatment to satisfactorily control the disease.
Furthermore, as recommended in the article by \citet{proust2008determinants}, we only considered the local relapses which took place three years or later after radiation, or within three years of EBRT when the last PSA value was $>2$ ng/ml.
PSA data were collected at regular visits, for a median number of 10 PSA measurements per patient. Note that this number includes only PSA data recorded between the end of EBRT and before the first event (first clinical recurrence, hormonal therapy, death or censorship).
Subjects with only one PSA measure were excluded, and subjects who had an event in the first year after EBRT  were excluded to prevent inclusion of patients who had substantial residual initial tumors.
As shown in Table~\ref{t:description_cohorts}, three baseline factors were considered: the pre-therapy level of PSA in the log scale (iPSA), the T-stage category which characterizes the tumour size (3 categories were considered: 2, 3--4 versus 1 in reference), and the Gleason score category which measures the aggressiveness of cancer cells (3 categories: 7, 8--10 versus 2--6 in reference). In the models, a cohort covariate was also considered coded as 1 for RTOG 9406 and $-1$ for BCCA. 
More details on the RTOG 9406 trial can be found in \citet{michalski2005toxicity}, and on the BCCA cohort in \citet{pickles2003evaluation}.

\begin{table}
\centering
\caption{Description of the two cohorts.}
\label{t:description_cohorts}
\begin{tabular*}{\columnwidth}{@{}l@{\extracolsep{\fill}}c@{\extracolsep{\fill}}c@{\extracolsep{\fill}}c@{}}
\hline
\multicolumn{1}{c}{Cohort} & RTOG 9406 & BCCA & Pooled \\
\hline
Study period & 1994--2013 & 1994--2012 &  \\
Number of patients & 629 & 845 & 1474 \\
Number of PSA measures per patient & 13 (4, 23) & 9 (3, 15) & 10 (3, 21) \\
iPSA$^\ast$ & 2.0 (1.0, 3.0) & 2.1 (0.6, 3.3) & 2.1 (0.8, 3.1) \\
Clinical T-stage & & & \\
\hspace{0.2cm} 1 & 355 (56.4\%) & 184 (21.8\%) & 539 (36.6\%) \\
\hspace{0.2cm} 2 & 261 (41.5\%) & 514 (60.8\%) & 775 (52.6\%) \\
\hspace{0.2cm} 3--4 & 13 (2.1\%) & 147 (17.4\%) & 160 (10.9\%) \\
Gleason score & & & \\
\hspace{0.2cm} 2--6 & 424 (67.4\%) & 605 (71.6\%) & 1029 (69.8\%) \\
\hspace{0.2cm} 7 & 167 (26.6\%) & 189 (22.4\%)  & 356 (24.2\%) \\
\hspace{0.2cm} 8--10 & 38 (6.0\%) & 51 (6.0\%) & 89 (6.0\%) \\
Mean time of first event$^\dagger$ & 9.8 (2.3, 15.9) & 7.7 (1.9, 14.1) & 8.2 (2.0, 15.0) \\
Mean time of last contact$^\ddagger$ & 11.6 (2.9, 16.7) & 9.0 (3.4, 14.8) & 9.7 (3.1, 15.9) \\
\hline
\multicolumn{4}{l}{Continuous data: Median (5th and 95th percentiles).}\\
\multicolumn{4}{l}{Categorical data: Amount (percentage).}\\
\multicolumn{4}{l}{Times are in years since the end of EBRT.}\\
\multicolumn{4}{l}{\up{$\ast$} Pre-therapy PSA value (ng/ml) in the $\log(. + 0.1)$ scale.}\\
\multicolumn{4}{l}{\up{$\dagger$} Minimum between the time of first transition and the time of censoring.}\\
\multicolumn{4}{l}{\up{$\ddagger$} Minimum between the time of death and the time of censorship.}
\end{tabular*}
\end{table}

\begin{figure}
\begin{center}
\centerline{\includegraphics[scale=1]{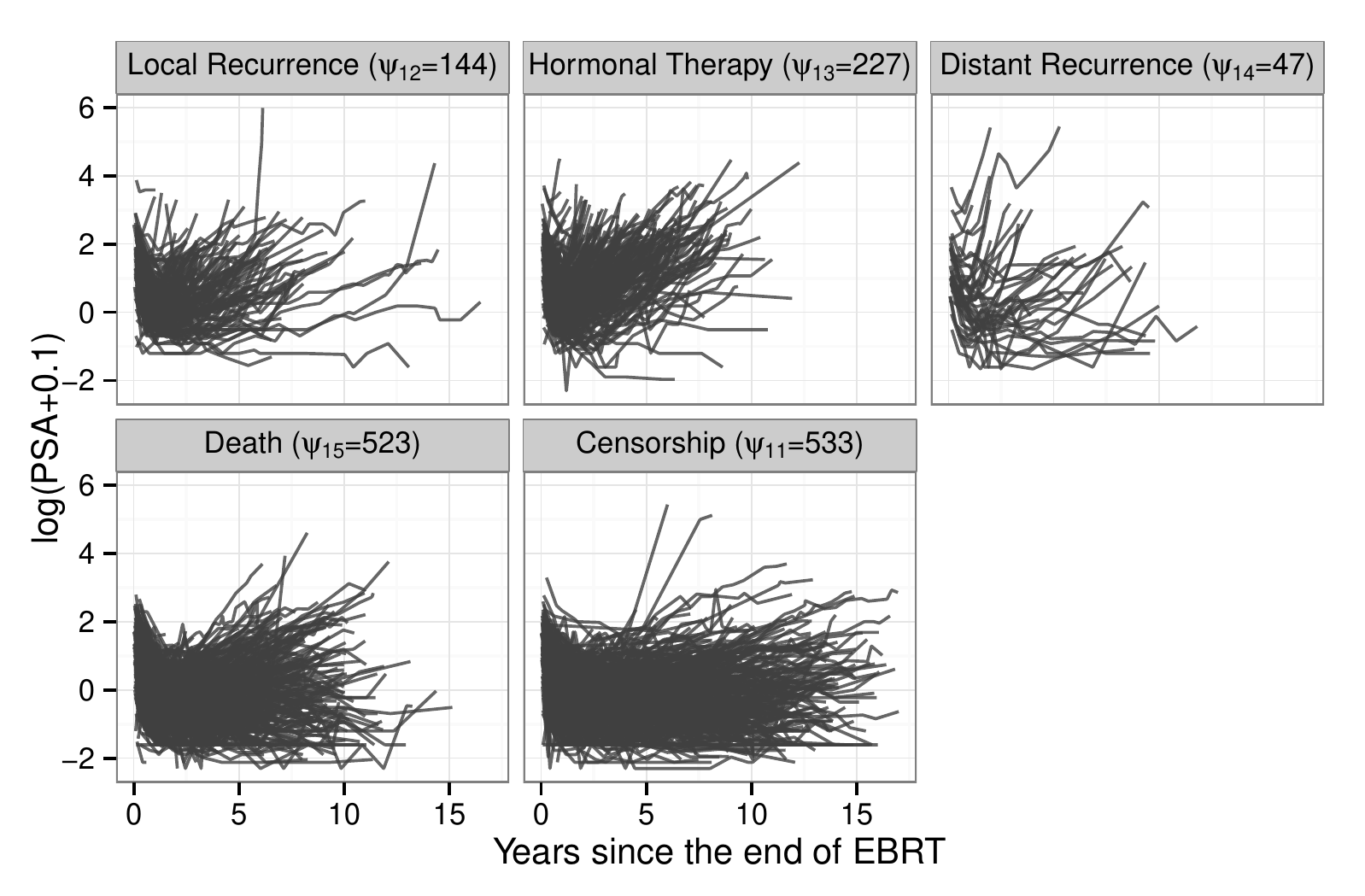}}
\end{center}
\caption{Individual trajectories of $\log \left( \textrm{PSA} + 0.1 \right)$ after the end of EBRT and according to the kind of first relapse in the two cohorts ($N = 1474$).}
\label{f:traj_PSA}
\end{figure}

The PSA individual trajectories collected between the end of EBRT and the occurrence of the first event are depicted in Figure~\ref{f:traj_PSA}.
Overall, this longitudinal process is biphasic, with a decrease in the level of PSA in the first years following the end of EBRT, and a subsequent stabilisation or linear rise thereafter. According to the type of first relapse, the biomarker's long-term increase may have different intensities (see ``Hormonal Therapy'' and ``Censorship'' for example).

The multi-state data are depicted through the transitions between the 5 states and the corresponding amount of observed direct transitions in Figure~\ref{f:observed_transitions}.
\begin{figure}
\centering
\begin{tikzpicture}[scale=1]
  \tikzstyle{noeud}=[minimum width=2cm,minimum height=1.33cm,line  width=1pt, rectangle, draw] 

  \node[draw,text width=2cm,minimum width=2cm,minimum height=1.33cm,text centered] (EBRT) at (0,-3.5) {\large End \\ EBRT};
    \draw (EBRT.base) node[below=18pt, right=22pt]{\footnotesize 0};
  \node[draw,text width=2cm,minimum width=2cm,minimum height=1.33cm,text centered] (RL) at (4,-3.5) {\large Local \\ Recurrence};
    \draw (RL.base) node[below=18pt, right=22pt]{\footnotesize 1};
  \node[draw,text width=2cm,minimum width=2cm,minimum height=1.33cm,text centered] (HT) at (7,0) {\large Hormonal \\ Therapy};
    \draw (HT.base) node[below=18pt, right=22pt]{\footnotesize 2};
  \node[draw,text width=2cm,minimum width=2cm,minimum height=1.33cm,text centered] (RD) at (7,-7) {\large Distant \\ Recurrence};
    \draw (RD.base) node[below=18pt, right=22pt]{\footnotesize 3};
  \node[draw,text width=2cm,minimum width=2cm,minimum height=1.33cm,text centered] (DC) at (10.5,-3.5) {\large Death};
    \draw (DC.base) node[below=10.5pt, right=22pt]{\footnotesize 4};

  \draw[->,>=latex,line width=1pt] (EBRT) -- (RL);
  \draw[->,>=latex,line width=1pt] (EBRT) -- (HT);
  \draw[->,>=latex,line width=1pt] (EBRT) -- (RD);
  \draw[->,>=latex,line width=1pt] (RL) -- (HT);
  \draw[->,>=latex,line width=1pt] (RL) -- (RD);
  \draw[->,>=latex,line width=1pt] (RL) -- (DC);
  \draw[->,>=latex,line width=1pt] (HT) -- (RD);
  \draw[->,>=latex,line width=1pt] (HT) -- (DC);
  \draw[->,>=latex,line width=1pt] (RD) -- (DC);
  \draw[->,>=latex,line width=1pt] (EBRT) ..controls +(4,-6.2) and +(0, -6.2).. (DC);
	
 \draw (2.45,-1.75) node {{\small $\lambda_{02}(t)$}};
 \draw (4.75,-1.75) node {{\small $\lambda_{12}(t)$}};
 \draw (7.65,-1.75) node {{\small $\lambda_{23}(t)$}};
 \draw (9.5,-1.75) node {{\small $\lambda_{24}(t)$}};
 
 \draw (2.4,-5.25) node {{\small $\lambda_{03}(t)$}};
 \draw (4.7,-5.25) node {{\small $\lambda_{13}(t)$}};
 \draw (9.5,-5.25) node {{\small $\lambda_{34}(t)$}};

 \draw (2,-3.2) node {{\small $\lambda_{01}(t)$}};
 \draw (6.1,-3.2) node {{\small $\lambda_{14}(t)$}};

 \draw (2.45,-7.25) node {{\small $\lambda_{04}(t)$}};
 
 \draw (-1,-10.2) node[right, text width=11.5cm] {$\boldsymbol{\Upsilon} = \begin{pmatrix} 533 & 144 & 227 & 47 & 523 \\ 0 & 20 & 90 & 10 & 24 \\ 0 & 0 & 106 & 33 & 178 \\ 0 & 0 & 0 & 13 & 77 \\ 0 & 0 & 0 & 0 & 802 \end{pmatrix}$} ;
\end{tikzpicture}
\caption{Multi-state representation of the clinical progressions in prostate cancer. Arrows indicate the directions of the possible transitions ($N=1474$). $\lambda_{hk}(t)$ characterizes the intensity of transition between states $h$ and $k$ at time $t$. The matrix $\boldsymbol{\Upsilon}$ has size $(5, 5)$ and is composed of elements $\Upsilon_{(h+1)(k+1)}$, where $\Upsilon_{(h+1)(k+1)}$ is the number of observed direct transitions $h \rightarrow k$. The diagonal elements $\Upsilon_{(h+1)(h+1)}$ denote the number of patients who were censored in state $h$. Note that the sum of elements of a row $(h+1)$ of $\boldsymbol{\Upsilon}$ corresponds to the number of patients who experienced the state $h$.}
\label{f:observed_transitions}
\end{figure}
From the end of EBRT (state 0), a patient can experience either a transition to a localized recurrence (state 1), an hormonal therapy (state 2), a distant recurrence (state 3) or death (absorbing state 4). After a localized recurrence (state 1), a patient may experience either a distant recurrence (state 3) or die (state 4) or initiate a HT (state 2). After initiation of HT, a patient may only experience a distant recurrence or die, and finally, after a distant recurrence, a patient may only die.
In total, 144 subjects had a local recurrence; 317 men initiated an hormonal therapy including 90 after a local recurrence; 90 men had a distant recurrence including 10 directly after a local recurrence and 33 after a HT initiation. In total, 802 patients died including 523 who did not have another recorded progression of the cancer before. Among the 672 men who were censored during the follow-up, 533 were censored before experiencing any clinical progression.

\subsection{Specification of the joint model}
\label{s:s:specification}
The joint multi-state model being a complex model, a step-by-step procedure was carried out to specify the joint model.
The specifications of the longitudinal and multi-state sub-models were based on two separate analyses, that is assuming independence between the two processes.
Covariate selection was made using uni- or multivariate Wald tests.

\subsubsection{Longitudinal sub-model specification}
\label{s:s::s:longit}
The biphasic shape of $\log$-PSA was described in a linear mixed model with two functions of time according to previous works \citep{proust2008determinants}: $f_1(t) = \left( 1 + t \right)^\alpha - 1$ and $f_2(t) = \left( t \right)^{ 1+\nu } / {\left( 1+t \right) ^{\nu}}$, where $\alpha$ and $\nu$ were estimated by profile likelihood ($\alpha = -1.2, \nu = 0$). Thus, these two functions depicted respectively the short term drop in the level of $\log$-PSA after EBRT and the long term linear increase of $\log$-PSA.
By denoting $Y_{ij} = \log \left( \textrm{PSA}_i(t_{ij}) + 0.1 \right)$ the $\log$-measure of PSA for the individual $i$ at time $t_{ij}$ --the natural logarithm transformation is performed to obtain a Gaussian shape for the longitudinal response-- the linear mixed sub-model took the form:
\begin{eqnarray*}
Y_{ij} & = &Y_{i}^{*}(t_{ij}) + \epsilon_{ij} \nonumber\\
	& =	&\left( \beta_0 + X^{L0~\top}_{i} \beta_{0,\texttt{cov}} + b_{i0} \right) + \nonumber\\
		&& \left( \beta_1 + X^{L1~\top}_{i} \beta_{1,\texttt{cov}} + b_{i1} \right) \times f_1(t_{ij}) + \nonumber\\
		&& \left( \beta_2 + X^{L2~\top}_{i} \beta_{2,\texttt{cov}} + b_{i2} \right) \times f_2(t_{ij}) + \epsilon_{ij} ,
\end{eqnarray*}
with $b_i = (b_{i0}, b_{i1}, b_{i2})^\top \sim \mathcal{N}\left( 0, D \right)$, $D$ unstructured, and $\epsilon_{i} = \left( \epsilon_{i1}, \ldots, \epsilon_{in_i} \right)^\top \sim \mathcal{N}(0, \sigma^2 I_{n_i})$.
The covariates $X^{L0}_{i}$, $X^{L1}_{i}$ and $X^{L2}_{i}$ were sub-vectors of the baseline prognostic factors obtained using a backward stepwise procedure.
For the sake of brevity, we will speak about PSA dynamics and biomarker's current level/slope when referring actually to respectively the dynamics of $\log(\textrm{PSA}+0.1)$ and the current level/slope of $Y^{*}_{i}(t)$.

\subsubsection{Multi-state sub-model specification}
\label{s:s:s:mstate}
In the multi-state sub-part, the determination of prognostic factors and proportionality between baseline intensities was also made by considering no link between the two processes ($\eta = 0$) and unspecified baseline intensities (i.e. using a standard semi-parametric multi-state model).
The full sub-model considered transition-specific baseline intensities and transition-specific effects of baseline prognostic factors.
To reduce the excessive number of parameters to estimate, a first step was to assume proportional baseline intensities for some transitions.
Clinically, it made sense to consider proportional baseline intensities for transitions leading to local recurrence or hormonal therapy: $\lambda_{01,0}(t) = \exp(\zeta_{02}) \lambda_{02,0}(t) = \exp(\zeta_{12}) \lambda_{12,0}(t)$; and for the transitions leading to distant recurrence: $\lambda_{03,0}(t) = \exp(\zeta_{13}) \lambda_{13,0}(t) = \exp(\zeta_{23}) \lambda_{23,0}(t)$. These assumptions were confirmed by the data.
We could not make the same assumption for all transitions leading to death because the proportional hazards assumption was not verified.
Instead, we chose $\lambda_{14,0}(t) = \exp(\zeta_{24}) \lambda_{24,0}(t)$ and $\lambda_{04,0}(t)$ was stratified on the cohort.
This procedure reduced the number of baseline intensities to six.
A second step consisted in selecting the prognostic factors.
Factors with an associated $\textrm{p-value} > 0.5$ were removed, and common covariate effects on several transitions were considered using multivariate Wald tests.
For example, the baseline T-stage category had the same effect on transition intensities $0 \rightarrow 1$, $0 \rightarrow 3$ and $2 \rightarrow 3$.
Finally, covariates with $\textrm{p-value} < 0.1$ were selected by using a backward stepwise procedure.

\subsubsection{Joint multi-state model specification}
\label{s:s:s:joint}
In the joint model, log baseline intensities approximated by linear combinations of cubic B-splines with three internal knots replaced the unspecified ones.
The dependence function $W_{hk,i}(b_i, t)$ was the same for all the transitions $h \rightarrow k$ and was determined using Wald tests. It resulted that the combination of the true current level and the true current slope of the biomarker fitted at best the relationship between PSA dynamics and the instantaneous risk to transit between health states.
The final step was to select the prognostic factors and the log-coefficients of proportionality between baseline intensities with $\textrm{p-value} < 0.1$ in the multi-state sub-part by using a backward stepwise procedure.
Thus, the multi-state sub-model was:
$$
\lambda^{i}_{hk}(t | b_i) = \lambda_{hk,0}(t)\exp \left({ X^{S~\top}_{hk,i} \gamma_{hk} + \begin{pmatrix} Y_{i}^{*}(t) \\ \partial Y_{i}^{*}(t) / \partial t \end{pmatrix}^{\top} \begin{pmatrix} \eta_{hk, \texttt{level}} \\ \eta_{hk,\texttt{slope}} \end{pmatrix} } \right) ,
$$
where the relations between $\lambda_{hk,0}(t)$ and the final $X^{S}_{hk,i}$, for $h,k \in \{0,\ldots,5\}$ are indicated in Section~\ref{s:s:s:mstate} and in Table~\ref{t:results}.
Note that the covariates that were removed of the joint model specification are not in Table~\ref{t:results}.

\subsection{Results}
\label{s:s:results}
The parameter estimates of the joint multi-state model are presented in Table~\ref{t:results}. These parameters were those selected according to the procedure described previously. The parameters of the baseline intensities are not shown here for clarity.

\begin{table}
\begin{center}
\caption{Parameter estimates, standard errors and $p$-values in the joint multi-state model on the pooled data ($N=1474$).}
\label{t:results}
\begin{tabular}{lrrrlrrr}
  \hline
  & \multicolumn{3}{c}{Longitudinal Process} & & \multicolumn{3}{c}{Multi-state Process}\\
  & Value & StdErr & $p$-value &   & Value & StdErr & $p$-value \\ 
  \hline
  $\beta_0$ & $-$0.26 & 0.06 & $<0.001$ & $\gamma_{02, \texttt{iPSA}}$ & 0.34 & 0.08 & $<0.001$ \\
  $\beta_{0,\texttt{iPSA}}$ & 0.80 & 0.03 & $<0.001$ & $\gamma_{04, \texttt{iPSA}}$ & 0.27 & 0.08 & $<0.001$ \\
  $\beta_{0,\texttt{cohort}}$ & $-$0.01 & 0.02 & $0.525$ & $\gamma_{(13,14,23,24,34), \texttt{iPSA}}$ & $-$0.24 & 0.08 & $0.002$ \\
  $\beta_1$ & 0.71 & 0.13 & $<0.001$ & $\gamma_{(01,03,23), \texttt{tstage2}}$ & 0.91 & 0.18 & $<0.001$ \\
  $\beta_{1,\texttt{iPSA}}$ & 0.89 & 0.06 & $<0.001$ & $\gamma_{(01,03,23), \texttt{tstage3-4}}$ & 0.73 & 0.23 & $0.002$ \\
  $\beta_{1,\texttt{tstage2}}$ & 0.38 & 0.08 & $<0.001$ & $\gamma_{(12,14,34), \texttt{tstage2}}$ & $-$0.04 & 0.25 & $0.865$ \\
  $\beta_{1,\texttt{tstage3-4}}$ & 0.48 & 0.13 & $<0.001$ & $\gamma_{(12,14,34), \texttt{tstage3-4}}$ & 0.38 & 0.30 & $0.204$ \\
  $\beta_{1,\texttt{cohort}}$ & $-$0.04 & 0.04 & $0.333$ & $\gamma_{(03,23) \texttt{gleason7}}$ & 0.97 & 0.24 & $<0.001$ \\
  $\beta_2$ & $-$0.24 & 0.04 & $<0.001$ & $\gamma_{(03,23) \texttt{gleason8-10}}$ & 0.09 & 0.43 & $0.827$ \\
  $\beta_{2,\texttt{iPSA}}$ & 0.19 & 0.02 & $<0.001$ & $\gamma_{(01,14,24,34), \texttt{cohort}}$ & $-$0.42 & 0.06 & $<0.001$ \\ 
  $\beta_{2,\texttt{tstage2}}$ & 0.14 & 0.02 & $<0.001$ & $\gamma_{(13,23), \texttt{cohort}}$ & 0.89 & 0.17 & $<0.001$ \\ 
  $\beta_{2,\texttt{tstage3-4}}$ & 0.26 & 0.04 & $<0.001$ & $\zeta_{(12,13)}$ & 4.18 & 0.38 & $<0.001$ \\
  $\beta_{2,\texttt{gleason7}}$ & 0.07 & 0.02 & $<0.001$ & $\zeta_{23}$ & 3.03 & 0.52 & $<0.001$ \\ 
  $\beta_{2,\texttt{gleason8-10}}$ & 0.22 & 0.04 & $<0.001$ & $\eta_{01, \texttt{level}}$ & 0.37 & 0.09 & $<0.001$ \\
  $\beta_{2,\texttt{cohort}}$ & $-$0.06 & 0.01 & $<0.001$ & $\eta_{02, \texttt{level}}$ & 0.51 & 0.07 & $<0.001$ \\
  $\log(\sigma)$ & $-$1.30 & 0.01 &  & $\eta_{03, \texttt{level}}$ & 0.45 & 0.11 & $<0.001$ \\  
   &  &  &  & $\eta_{04, \texttt{level}}$ & $-$0.17 & 0.05 & $0.001$ \\
  $D_{11}$ & 0.37 & 0.02 & & $\eta_{12, \texttt{level}}$ & $-$0.16 & 0.10 & $0.110$ \\
  $D_{12}$ & 0.01 & 0.01 & & $\eta_{13, \texttt{level}}$ & $-$0.41 & 0.20 & $0.042$ \\
  $D_{13}$ & 0.35 & 0.03 & & $\eta_{14, \texttt{level}}$ & 0.10 & 0.14 & $0.487$ \\
  $D_{22}$ & 0.14 & 0.01 & & $\eta_{23, \texttt{level}}$ & $-$0.15 & 0.09 & $0.120$ \\ 
  $D_{23}$ & 0.25 & 0.02 & & $\eta_{24, \texttt{level}}$ & 0.04 & 0.05 & $0.412$ \\ 
  $D_{33}$ & 1.68 & 0.09 & & $\eta_{34, \texttt{level}}$ & 0.04 & 0.08 & $0.609$ \\
   &  &  &  & $\eta_{01, \texttt{slope}}$ & 2.54 & 0.31 & $<0.001$ \\ 
   &  &  &  & $\eta_{02, \texttt{slope}}$ & 3.04 & 0.25 & $<0.001$ \\ 
   &  &  &  & $\eta_{03, \texttt{slope}}$ & 2.43 & 0.49 & $<0.001$ \\ 
   &  &  &  & $\eta_{04, \texttt{slope}}$ & 1.03 & 0.32 & $0.001$ \\ 
   &  &  &  & $\eta_{12, \texttt{slope}}$ & 2.01 & 0.61 & $0.001$ \\ 
   &  &  &  & $\eta_{13, \texttt{slope}}$ & 3.18 & 0.80 & $<0.001$ \\ 
   &  &  &  & $\eta_{14, \texttt{slope}}$ & $-$0.20 & 1.27 & $0.873$ \\ 
   &  &  &  & $\eta_{23, \texttt{slope}}$ & 0.97 & 0.67 & $0.150$ \\ 
   &  &  &  & $\eta_{24, \texttt{slope}}$ & 0.29 & 0.52 & $0.583$ \\ 
   &  &  &  & $\eta_{34, \texttt{slope}}$ & $-$0.79 & 0.78 & $0.313$ \\ 
   \hline \\
\end{tabular}
\end{center}
$D_{ij}$ denotes the $ij$-element of the covariance matrix for the random effects. $\gamma_{(hk,h'k'), \texttt{X}}$ denotes the effect of the covariate $\texttt{X}$ on the intensities of transitions $h \rightarrow k$ and $h' \rightarrow k'$, i.e.$\gamma_{(hk,h'k'), \texttt{X}} = \gamma_{hk, \texttt{X}} = \gamma_{h'k', \texttt{X}}$. In the same idea, it is used $\zeta_{(12,13)} = \zeta_{12} = \zeta_{13}$.
\end{table}
The estimated regression parameters in the longitudinal sub-part confirmed that pre-treatment PSA level was associated with the initial PSA level and the biphasic PSA trajectory, T-stage value was associated both with the short term and the long term dynamics while Gleason score was only associated with the long term trajectory.
Higher values of these covariates measured at baseline corresponded to higher long term PSA levels.
The cohort effect indicated a significant difference between the two cohorts only for the long term PSA evolution, with a greater long term increase of PSA in Vancouver.

For the multi-state process, the prognostic factors showed that an advanced initial stage was not always associated with the intensities of transitions between health states after adjustment for the PSA dynamics. In particular, the Gleason score had significant effects on only two transition intensities.
Moreover, we found that having a high PSA value at baseline was significantly associated with a higher instantaneous risk to directly experience hormonal therapy initiation or death after EBRT, but reduced the intensities of transitions leading to distant recurrence or death after a previous event.
A poor (i.e. higher) T-stage category at baseline had globally a deleterious effect on the clinical endpoints.
For the transitions leading to distant recurrence after EBRT or after hormonal therapy, a patient with a Gleason score of 7 at baseline had a $2.65 = \exp(0.973)$ (95\% CI = 1.64--4.28) higher hazard to transit than a patient with a Gleason score $<7$.
The cohort was significantly associated with the intensities of transitions leading to death after clinical recurrence or hormonal therapy --and the direct transition leading to local recurrence after EBRT.
The instantaneous risk to experience these transitions was higher in BCCA.
For the direct transitions leading to distant recurrence after local recurrence or hormonal therapy, the cohort effect was also significant, with higher intensities in RTOG 9406.

Regarding the association parameters between PSA dynamics (current level and current slope) and clinical progressions, there were highly significant deleterious effects of the PSA dynamics from the initial state on the intensities of transitions leading to all the types of progression (local recurrence, hormonal therapy or distant recurrence).
For example, after adjustments for covariates and for the true slope of the biomarker, an increase of one unit of the true biomarker's level (log PSA without error measurement) induced a $1.45 = \exp(0.372)$ (95\% CI = 1.23--1.72) higher risk to experience a local recurrence.
These results were expected: in patients with localized prostate cancer and treated by radiotherapy, a persistently high PSA level or/and a strong increase of PSA leads to higher hazard to experience a clinical recurrence or an additional therapy.
In contrast, for the direct transition leading to death after radiotherapy, we found a deleterious effect of the current slope and a protective effect of the current level of the biomarker: at a given moment in the initial state, for two patients with the same baseline characteristics and the same slope of log PSA, the one with higher PSA value will be less likely to directly die.
In this studied population, an important cause of death is induced by comorbidities, most of death from prostate cancer experienced a documented disease progression before.
From the local recurrence, there was large deleterious effect of the current slope of the biomarker for the intensities of transitions leading to the hormonal therapy or the distant recurrence, and there was a borderline significant protective effect of the current level for the intensity of transition leading to the distant recurrence.
From the hormonal therapy or the distant recurrence, there was no significant effect of the PSA dynamics on the hazard to change state.
This was also clinically sensible, as it reflects that progression in these advanced stages is no more linked to PSA increase. 
In practice, criteria other than PSA are considered specifically in this phase of the disease, such as the PCWG2 criteria \citep{scher2008design}. Moreover, deaths in patients with hormonal therapy might be explained by cardiac toxicity due to HT.

\subsection{Diagnostics}
\label{s:s:diagnostics}

\begin{figure}
\begin{center}
\begin{tabular}{cc}
\subfloat[Conditional standardized residuals versus the fitted values. The black curve indicates the mean of conditional residuals according to the fitted values, using a locally weighted polynomial regression.]{\includegraphics[width = 0.46\linewidth]{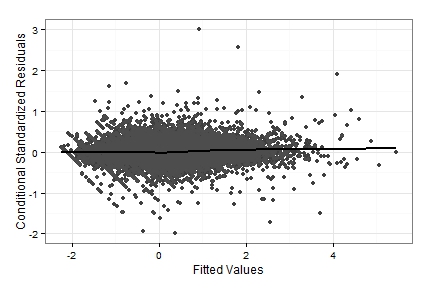}} & 
\subfloat[Observed and predicted values of the biomarker. 95\% confidence intervals of the observed values are connected in grey.]{\includegraphics[width = 0.46\linewidth]{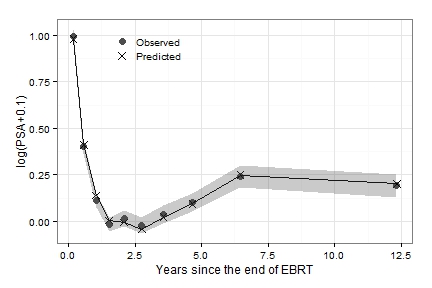}} \\
\multicolumn{2}{c}{\subfloat[Predicted transition probabilities from the joint multi-state model and non-parametric probability transitions.]{\includegraphics[width = 1\linewidth]{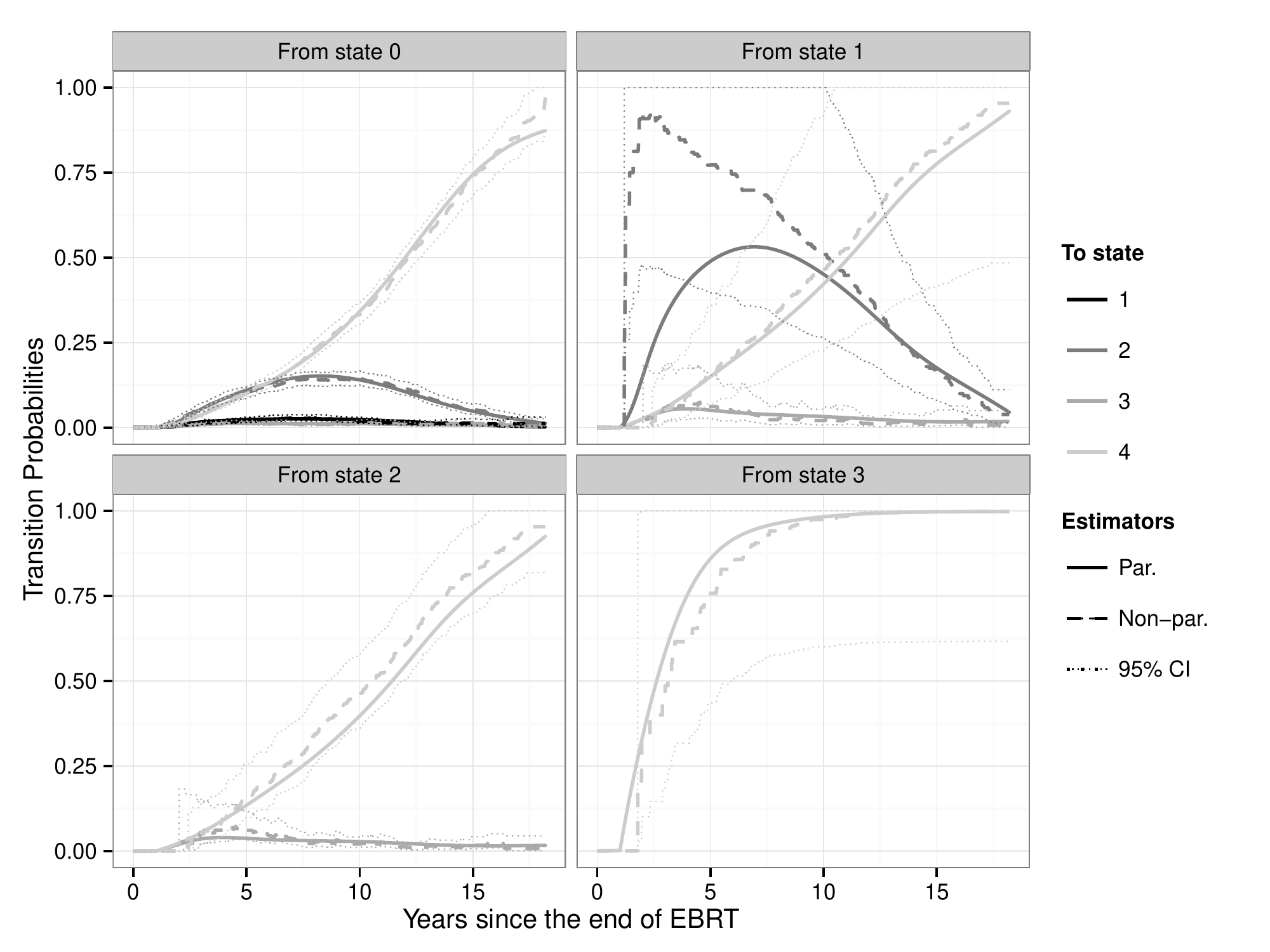}}}
\end{tabular}
\vspace{1cm}
\end{center}
\caption{Goodness-of-fit plots for the longitudinal process (a,b) and the multi-state process (c).}
\label{f:GOF}
\end{figure}

The parameter estimates of the joint multi-state model were validated by several graphical tools presented in Figure~\ref{f:GOF}. For the longitudinal sub-model, the plotted standardized conditional residuals versus fitted values of the biomarker confirmed the homoscedasticity of the conditional errors.
Subject-specific predictions were also compared to observations by plotting the average values by time intervals based on the deciles of the observation times. 95\% confidence intervals of the observed values were added and confirmed the very good fit of the model concerning the longitudinal data.
For the multi-state sub-part, we focused on $\mathbf{P}(0,t) = \lbrace P_{hk}(0,t) \rbrace$ the matrix of transition probabilities between the times $0$ and $t$.
We compared our parametric estimator (obtained through the average of the predicted individual transition probabilities from the joint multi-state model) to the Aalen-Johansen estimator (non-parametric estimator of the transition probabilities), both using product integrals. This comparison is fully discussed and detailed in Appendix B. 
These comparisons showed the overall good performances of the joint multi-state model in terms of fit of transition probabilities, with the exception for the transition $1 \rightarrow 2$ for which the immediate pike after EBRT could not be correctly captured by splines.

\section{Discussion}
\label{s:discussion}

The joint model for the longitudinal biomarker PSA and multi-state clinical progression data provides a full model of prostate cancer progression which takes into account both classical prognostic factors and the dynamics of the PSA, in order to study factors that influence the transition intensities between clinical health states.
The implementation is easy and relies on the \texttt{mstate} and \texttt{JM} packages.
The multi-state data are prepared through the \texttt{mstate} package, and a slightly modified \texttt{jointModel()} function carries out the estimation procedure.
The estimation program has been validated by simulations, with very good performances.
It underlined however some bias in the estimates when too few quadrature points (3) were used. In the absence of specific validation, we thus recommend at least 9 Gauss-Hermite quadrature points in the pseudo-adaptative numerical integration to approximate at best the integral over the random effects.
Goodness-of-fit of the model can be assessed using diagnostic graphical tools whose methodology was detailed in Appendix B.

The application confirmed that the PSA dynamics strongly impacted the instantaneous risk to experience a clinical recurrence or hormonal therapy after the end of radiotherapy. Moreover, the current slope of the biomarker had highly significant deleterious effect on the hazard to transit from local recurrence to hormonal therapy or distant recurrence.
Conversely, extrapolating the biomarker's dynamics did not impact anymore the transition intensities from the hormonal therapy or the distant recurrence states. This expresses that in the advanced cancers, the PSA --and especially the collected measures prior the first event-- is no more of importance. In these situations, other criteria have to be monitored.

Previous works in prostate cancer had found a strong association between slope of log-PSA and any clinical recurrence (see \citet{sene2014shared, taylor2013real}), by considering all the recurrences in a composite event and the hormonal therapy as a time-dependent covariate.
The limit of these approaches was that in practice the type of progression is of major importance as the care greatly depends on the type of risk the patient has. The joint multi-state model formalizes this need.
In the same way as it was done with a single event (see \citet{proust2009development, rizopoulos2011dynamic}), individualized dynamic predictions of each type of progression could be derived from this model in order to precisely quantify the risk of each type of progression according to the PSA history. 
For example, the cumulative probability for subject $i$ to reach the state $k$ between the times $s$ and $t$, $s \leq t$, given he was in state $h$ at time $s$, could be expressed as: $F^{i}_{hk}(s,t) = \int_{s}^{t}{ \Pr( E_i(u) = k | E_i(s) = h, Y^{(s)}_{i}, X^{L~(s)}_{i}, X^{S}_{i} )} \, \mathrm du$, with $Y^{(s)}_{i}$ the history (i.e. collected measures) of the marker up to time $s$, $X^{L~(s)}_{i}$ the history of the longitudinal sub-part covariates until time $s$, and $X^{S}_{i} = \{ X^{S}_{hk, i} \}$ the matrix containing the prognostic factors for all the state transitions.

In this article, we assumed a continuous and Markov multi-state process.
Depending on the topic, a semi-Markov process, which considers the spent time in the current state, could be defined as well. In dementia for example, the multi-state process might consider three states: healthy, demented and dead, and it would be necessary to consider the time spent in the demented state before death \citep{commenges2007choice}. The joint multi-state model and its associated implemented function handle for semi-Markov.

In summary, we introduce here a first joint model for longitudinal and multi-state clinical progression data.
We showed that this model can easily be implemented under R and can be applied in practice through an example, the prostate cancer progression, which is one of many biomedical areas in which such data are collected.
This model that captures the complete information about the progression opens to much more precise knowledge of diseases and specific dynamic predictions.

\section{Appendix}

\subsection*{Appendix A: Simulation results}
\label{ss:AppA}
In Section 4 in the article, the simulations were performed on 1500 subjects and showed that 9 Gauss-Hermite quadrature points should be used in this case.
A complementary simulation study was based on the first 500 and 1000 individuals of each of the 500 replicates and used 9 Gauss-Hermite quadrature points.
With 500 and 1000 subjects, we respectively observed 167 and 333 transitions $0 \rightarrow 1$, 103 and 205 transitions $0 \rightarrow 2$, 96 and 191 transitions $1 \rightarrow 2$.
The results are presented in Table~\ref{t:simulation_results}.

\begin{table}
\centering
\begin{small}
\caption[Simulation results on the 500 and 1000 first individuals of the 500 replicates and using 9 Gauss-Hermite quadrature points. For each scenario, the statistics depicted are, from left to right: mean,  mean standard error, standard deviation, relative bias, root-mean-square-error and coverage rate.]{Simulation results on the 500 and 1000 first individuals of the 500 replicates and using 9 Gauss-Hermite quadrature points. For each scenario, the statistics depicted are, from left to right: mean,  mean standard error, standard deviation, relative bias (in percentage) and coverage rate (in percentage).}
\label{t:simulation_results}
\begin{tabular}{lrrrrrrrrrrr}
  \hline
 & & \multicolumn{5}{c}{500 subjects} & \multicolumn{5}{c}{1000 subjects} \\
 & & \multicolumn{5}{c}{\hrulefill} & \multicolumn{5}{c}{\hrulefill} \\
 & True & \multirow{2}{*}{Mean} & \multirow{2}{*}{$\overline{\textrm{StdErr}}$} & \multirow{2}{*}{StdDev} & Rel. & Cov. & \multirow{2}{*}{Mean} & \multirow{2}{*}{$\overline{\textrm{StdErr}}$} & \multirow{2}{*}{StdDev} & Rel. & Cov. \\
  & value & & & & bias & rate & & & & bias & rate \\
  \hline
\multicolumn{2}{l}{\hspace{-0.35cm} \textit{Longitudinal process}} &&&&&&&&&& \\
$\beta_0$ & $-$0.793 & $-$0.797 & 0.085 & 0.086 & 0.4 & 94.8 & $-$0.798 & 0.060 & 0.061 & 0.6 & 94.4 \\
  $\beta_{0, X}$ & 0.543 & 0.544 & 0.039 & 0.040 & 0.2 & 94.8 & 0.545 & 0.028 & 0.029 & 0.3 & 95.8 \\
  $\beta_1$ & $-$0.096 &  -0.096 & 0.036 & 0.036 & $-$0.2 & 96.0 & $-$0.096 & 0.026 & 0.025 & 0.1 & 95.2 \\
  $\beta_{1, X}$ & 0.027 & 0.027 & 0.017 & 0.017 & $-$0.1 & 95.6 & 0.027 & 0.012 & 0.012 & $-$0.1 & 95.2 \\
  $\log(\sigma)$ & $-$0.737 & $-$0.737 & 0.007 & 0.019 & $-$0.1 & 96.0 & $-$0.737 & 0.005 & 0.005 & -0.0 & 93.0 \\
\multicolumn{2}{l}{\hspace{-0.35cm} \textit{Multi-state process}} &&&&&&&&&& \\
  $\gamma_{01, X}$ & 0.281 & 0.298 & 0.136 & 0.136 & 5.9 & 95.0 & 0.296 & 0.095 & 0.094 & 5.2 & 95.2 \\
  $\gamma_{02, X}$ & 0.023 & 0.010 & 0.155 & 0.161 & $-$57.9 & 93.0 & 0.028 & 0.108 & 0.112 & 21.6 & 93.8 \\
  $\gamma_{12, X}$ & $-$0.169 & $-$0.162 & 0.171 & 0.176 & $-$4.3 & 94.6 & $-$0.164 & 0.117 & 0.125 & $-$3.1 & 94.8 \\
  $\eta_{01, \texttt{level}}$ & 0.925 & 0.927 & 0.130 & 0.144 & 0.2 & 93.0 & 0.910 & 0.090 & 0.091 & $-$1.6 & 94.6 \\
  $\eta_{02, \texttt{level}}$ & 0.297 & 0.298 & 0.113 & 0.109 & 0.6 & 96.0 & 0.292 & 0.079 & 0.075 & $-$1.5 & 95.6 \\
  $\eta_{12, \texttt{level}}$ & 0.071 & 0.071 & 0.135 & 0.141 & 0.3 & 94.4 & 0.072 & 0.092 & 0.093 & 0.4 & 95.4 \\
  $\eta_{01, \texttt{slope}}$ & 1.344 & 1.514 & 0.773 & 0.827 & 12.6 & 93.6 & 1.531 & 0.541 & 0.524 & 13.9 & 95.4 \\
  $\eta_{02, \texttt{slope}}$ & $-$1.096 & -1.071 & 1.125 & 1.119 & $-$2.3 & 97.6 & $-$1.041 & 0.788 & 0.737 & $-$5.0 & 95.6 \\
  $\eta_{12, \texttt{slope}}$ & 0.009 & 0.086 & 1.425 & 1.442 & 853.3 & 96.4 & 0.036 & 0.976 & 1.007 & 298.2 & 94.6 \\
\multicolumn{2}{l}{\hspace{-0.35cm} \textit{Random effects}} &&&&&&&&&& \\
  $D_{11}$ & 0.349 & 0.346 & 0.025 & 0.025 & $-$0.8 & 95.0 & 0.347 & 0.017 & 0.017 & $-$0.6 & 95.4 \\
  $D_{12}$ & $-$0.041 & $-$0.042 & 0.008 & 0.008 & 0.3 & 95.0 & $-$0.042 & 0.006 & 0.006 & 0.5 & 95.8 \\
  $D_{22}$ & 0.062 & 0.062 & 0.004 & 0.005 & $-$0.5 & 93.8 & 0.062 & 0.003 & 0.003 & $-$0.2 & 92.6 \\
   \hline \\
\end{tabular}
\end{small}
\end{table}
Overall, these results confirmed the good performances of the implemented function. The coverage rates were close to 95\% and the relative bias were low, except for $\gamma_{02,X}$ and $\eta_{12,\texttt{slope}}$, certainly due to the required accuracy.

\subsection*{Appendix B: Parametric versus non-parametric transition probabilities}
\label{ss:AppB}

To assess the results obtained in our application (Section 5 in the article), we compared the parametric estimator of the transition probabilities to a non-parametric estimator. The following results are based on the book of \citet{andersenstatistical}. 

\subsubsection*{Preliminaries}
\label{sss:Preliminaries}

Consider the multi-state process $E = \lbrace E(t), t \geq 0 \rbrace$ with values in the finite space $S = \lbrace 0,1,\ldots,M\rbrace$ and where $E(t)$ denotes the state occupied by an individual at time $t$. We assume that E is a non-homogeneous Markov process, with left truncation and right censoring.
In the following, we will consider that all the introduced multi-state processes guarantee the above properties.
The intensity of transition from state $h \in S$ to state $k \in S$ at time $t$ is defined as $\lambda_{hk}(t) = \lim_{\mathrm dt \rightarrow 0} \dfrac{\Pr(E(t + \mathrm dt)=k | E(t)=h)}{\mathrm dt}$ and we write $\boldsymbol{\lambda}(s,t) = \{ \lambda_{hk}(s,t) \}$ the $(M+1) \times (M+1)$ matrix of transition intensities.
The matrix of cumulative transition intensities is noted $\mathbf{\Lambda}(t),$ composed of non-diagonal elements $\Lambda_{hk}(t) = \int_{0}^{t} \lambda_{hk}(u) \, \mathrm du, \forall h \neq k,$ and diagonal elements $\Lambda_{hh}(t) = - \sum_{k \neq h}\Lambda_{hk}(t)$.
Let us consider the transition probability $P_{hk}(s,t) = \Pr(E(t) = k | E(s) = h)$, with $s \leq t$, which is the probability that a subject in state $h$ at time $s$ occupies the state $k$ at a later time $t$.
We call $\mathbf{P}(s,t) = \{ P_{hk}(s,t) \}$ the matrix of transition probabilities, which satisfies the Chapman-Kolmogorov equation: 
\begin{eqnarray}
\mathbf{P}(s,t) = \mathbf{P}(s,u)\mathbf{P}(u,t), \textrm{ with } 0 \leq s \leq u \leq t.
\end{eqnarray}
Thus is deduced that $\mathbf{P}(s,t)$ is the unique solution of the Kolmogorov forward differential equations:
\begin{eqnarray}
\label{e:kolmogorov}
\begin{array}{c}
\mathbf{P}(s,s) = \mathbf{I},\\
\dfrac{\partial}{\partial t} \mathbf{P}(s,t) = \mathbf{P}(s,t) \boldsymbol{\lambda}(t).
\end{array}
\end{eqnarray}

\subsubsection*{Non-parametric estimator}
\label{sss:Non-par}

Let $N_{hk}(t)$ be the number of direct observed transitions from state $h$ to state $k$ up to time $t$, and $Y_h(t)$ the number of individuals in state $h$ just before time $t$.
The non-parametric estimator of the cumulative intensities, called $\mathbf{\Lambda}^{*}(t)$ has elements $\Lambda^{*}_{hk}(t)$ estimated through the Nelson-Aalen estimator:
\begin{eqnarray}
\Lambda^{*}_{hk}(t) = \displaystyle{\int_{0}^{t}} \dfrac{\mathrm dN_{hk}(u)}{Y_h(u)}, h \neq k,
\end{eqnarray}
and $\Lambda^{*}_{hh}(t) = - \sum_{k \neq h} \Lambda^{*}_{hk}(t)$.
The solution to the Kolmogorov equations (\ref{e:kolmogorov}) permits to express the non-parametric estimate of the transition probabilities $\mathbf{\widehat{P}}^{*}(s,t)$ though the product-integral:
\begin{eqnarray}
\label{e:non-par}
\mathbf{\widehat{P}}^{*}(s,t) = \Prodi_{(s,t]}{\left(\boldsymbol{\mathrm{I}} + \mathrm d\mathbf{\widehat{\Lambda}}^{*}(u)\right)}.
\end{eqnarray}
where $\mathbf{\widehat{\Lambda}}^{*}(u)$ is the non-parametric estimate of the cumulative transition intensities at time $u$, with $\mathrm d\widehat{\Lambda}_{hh}^{*}(u) \geq -1$ for all $u$, and $\boldsymbol{\mathrm{I}}$ is the $(M+1) \times (M+1)$ identity matrix.
This estimated matrix of transition probabilities is called Aalen-Johansen estimator.

Let $s < T_{1}^{*} < \ldots < T_{m^*}^{*} \leq t$ be the ordained times of observed direct transitions between $s$ and $t$ for all individuals. From (\ref{e:non-par}), it can be deduced:
\begin{eqnarray}
\mathbf{\widehat{P}}^{*}(s,t) = \prod_{l=1}^{m^*}{\left(\boldsymbol{\mathrm{I}} + \Delta\mathbf{\widehat{\Lambda}}^{*}(T_{l}^{*})\right)}.
\end{eqnarray}
where $\Delta\mathbf{\widehat{\Lambda}}^{*}(T_{l}^{*}) = \mathbf{\widehat{\Lambda}}^{*}(T_{l}^{*}) - \mathbf{\widehat{\Lambda}}^{*}(T_{l-1}^{*})$ and $\Delta \widehat{\Lambda}_{hh}^{*}(T_{l}^{*}) \geq -1$ for all $T_{l}^{*}$.

In our application, we applied:
$$
\boldsymbol{\mathrm{I}} + \Delta \mathbf{\widehat{\Lambda}}^{*}(T_{l}^{*}) = 
\begin{pmatrix} 
1 - \dfrac{\Delta N_{0.}(T_{l}^{*})}{Y_0(T_{l}^{*})} & \dfrac{\Delta N_{01}(T_{l}^{*})}{Y_0(T_{l}^{*})} & \dfrac{\Delta N_{02}(T_{l}^{*})}{Y_0(T_{l}^{*})} & \dfrac{\Delta N_{03}(T_{l}^{*})}{Y_0(T_{l}^{*})} & \dfrac{\Delta N_{04}(T_{l}^{*})}{Y_0(T_{l}^{*})} \\ 
0 & 1 - \dfrac{\Delta N_{1.}(T_{l}^{*})}{Y_1(T_{l}^{*})} & \dfrac{\Delta N_{12}(T_{l}^{*})}{Y_1(T_{l}^{*})} & \dfrac{\Delta N_{13}(T_{l}^{*})}{Y_1(T_{l}^{*})} & \dfrac{\Delta N_{14}(T_{l}^{*})}{Y_1(T_{l}^{*})} \\
0 & 0 & 1 - \dfrac{\Delta N_{2.}(T_{l}^{*})}{Y_2(T_{l}^{*})} & \dfrac{\Delta N_{23}(T_{l}^{*})}{Y_2(T_{l}^{*})} & \dfrac{\Delta N_{24}(T_{l}^{*})}{Y_2(T_{l}^{*})} \\
0 & 0 & 0 & 1 - \dfrac{\Delta N_{3.}(T_{l}^{*})}{Y_3(T_{l}^{*})} & \dfrac{\Delta N_{34}(T_{l}^{*})}{Y_3(T_{l}^{*})} \\
0 & 0 & 0 & 0 & 1
\end{pmatrix},
$$
where $N_{h.}(T_{l}^{*}) = \sum_{k \neq h} N_{hk}(T_{l}^{*})$ and $\Delta N_{hk}(T_{l}^{*}) = N_{hk}(T_{l}^{*}) - N_{hk}(T_{l-1}^{*})$.

\subsubsection*{Parametric estimator}
\label{sss:Par}

For each patient $i \in \{1,\ldots,N\}$, the observed multi-state process is $\lbrace E_i(t), t \geq 0 \rbrace$, where $E_i(t)$ denotes the state occupied by subject $i$ at time $t$ and takes values in the finite space $S = \lbrace 0,1,\ldots,M\rbrace$. In our joint multi-state model, we were interested in the transitions intensities:
\begin{eqnarray}
\lambda^{i}_{hk}(t | b_i) = \lim_{\mathrm dt \rightarrow 0} \dfrac{\Pr(E_i(t + \mathrm dt)=k | E_i(t)=h; b_i)}{\mathrm dt},
\end{eqnarray}
which share the random effects $b_i$ with the longitudinal sub-model.
Based on the \textit{Preliminaries} paragraph, let $\mathbf{\Lambda}^{i}(t | b_i) = \{ \Lambda^{i}_{hk}(t | b_i) \}$ be the parametric matrix of cumulative intensities for subject $i$, and $\mathbf{P}^{i}_{hk}(s,t | b_i) = \{ P^{i}_{hk}(s,t | b_i) \}$ be the parametric matrix of transition probabilities.
From the individual covariates measured at baseline $X^{S}_{hk,i}$ and the parameters estimated by maximum likelihood $\hat{\theta}$, we computed for each subject $i$ the individual predictions of the transition intensities $\widehat{\lambda}^{i}_{hk}(t | \hat{\theta})$ and deduced the individual predictions of the cumulative intensities $\widehat{\Lambda}^{i}_{hk}(t | \hat{\theta})$.
To obtain the parametric estimator of the individual transition probabilities $\mathbf{\widehat{P}}^{i}(s,t | \hat{\theta})$, we could use the relations:
\begin{eqnarray}
\label{e:par}
\begin{array}{c}
\widehat{P}^{i}_{hh}(s,t | \hat{\theta}) = \exp \left(\widehat{\Lambda}^{i}_{hh}(t | \hat{\theta}) - \widehat{\Lambda}^{i}_{hh}(s | \hat{\theta})\right), \\
\widehat{P}^{i}_{hk}(s,t | \hat{\theta}) = \displaystyle{\int_{s}^{t}} \widehat{P}^{i}_{hh}(s,u | \hat{\theta}) \widehat{\lambda}^{i}_{hk}(u | \hat{\theta}) \widehat{P}^{i}_{kk}(u,t | \hat{\theta}) \mathrm du, h \neq k.
\end{array}
\end{eqnarray}
In practice, when the state space $S$ is large, these integrals become too complex numerically. In the application we therefore preferred to calculate $\mathbf{\widehat{P}}^{i}(s,t | \hat{\theta})$ through the product-integral:
\begin{eqnarray}
\label{e:par}
\mathbf{\widehat{P}}^{i}(s,t | \hat{\theta}) = \Prodi_{(s,t]}{\left(\boldsymbol{\mathrm{I}} + \mathrm d\mathbf{\widehat{\Lambda}}^{i}(u)\right)},
\end{eqnarray}
with $\mathrm d \widehat{\Lambda}^{i}_{hh}(u) \geq -1$ for all $u$.
The matrix of transition probabilities $\mathbf{\widehat{P}}(s,t)$ was then deduced by averaging over the N individual predictions:
\begin{eqnarray}
\mathbf{\widehat{P}}(s,t | \hat{\theta}) = \dfrac{1}{N} \sum_{i=1}^{N} \mathbf{\widehat{P}}^i(s,t | \hat{\theta}).
\end{eqnarray}

\subsubsection*{Covariance estimation of the non-parametric estimator}
\label{sss:Cov}

The covariance matrix of the Aalen-Johansen estimator of transition probabilities can be estimated by the Greenwood-type estimator:
\begin{eqnarray}
\label{e:cov}
\widehat{\mathrm{cov}}(\mathbf{\widehat{P}}^{*}(s,t)) = \displaystyle{\int_{s}^{t}} \mathbf{\widehat{P}}^{*}(u,t)^\top \otimes \mathbf{\widehat{P}}^{*}(s,u-) \widehat{\mathrm{cov}}(\mathrm d\mathbf{\widehat{\Lambda}}^{*}(u)) \mathbf{\widehat{P}}^{*}(u,t)) \otimes \mathbf{\widehat{P}}^{*}(s,u-)^\top ,
\end{eqnarray}
where $\otimes$ denotes the Kronecker product and $^\top$ denotes the vector transpose.

\citet{andersenstatistical} simplified the computations in (\ref{e:cov}) using the recursion formula:
\begin{eqnarray}
\widehat{\mathrm{cov}}(\mathbf{\widehat{P}}^{*}(s,t)) & = & \{ (\boldsymbol{\mathrm{I}} + \Delta \mathbf{\widehat{\Lambda}}^{*}(t))^\top \otimes \mathbf{I}\} \widehat{\mathrm{cov}}(\mathbf{\widehat{P}}^{*}(s,t-)) \{(\boldsymbol{\mathrm{I}} + \Delta \mathbf{\widehat{\Lambda}}^{*}(t)) \otimes \mathbf{I}\} + \nonumber\\
&& \{ \mathbf{I} \otimes \mathbf{\widehat{P}}^{*}(s,t-)\} \widehat{\mathrm{cov}}(\Delta \mathbf{\widehat{\Lambda}}^{*}(t)) \{ \mathbf{I} \otimes \mathbf{\widehat{P}}^{*}(s,t-) \},
\end{eqnarray}
where
\begin{eqnarray*}
\widehat{\mathrm{cov}}(\Delta \widehat{\Lambda}_{hk}^{*}(t), \Delta \widehat{\Lambda}_{h^\prime k^\prime}^{*}(t)) & = & 
\left\{
\begin{array}{l}
\dfrac{(Y_h(t) - \Delta N_{h.}(t))\Delta N_{h.}(t)}{Y_{h}(t)^3}, \textrm{ for } h=k=h^\prime=k^\prime, \nonumber \\
- \dfrac{(Y_h(t) - \Delta N_{h.}(t))\Delta N_{hk^\prime}(t)}{Y_{h}(t)^3}, \textrm{ for } h=k=h^\prime \neq k^\prime, \nonumber \\
- \dfrac{(\delta_{kk^\prime} Y_h(t) - \Delta N_{hk}(t))\Delta N_{hk^\prime}(t)}{Y_{h}(t)^3}, \textrm{ for } h=h^\prime, h \neq k, h \neq k^\prime, \nonumber \\
0, \textrm{ for } h \neq h^\prime, \nonumber
\end{array}
\right.
\end{eqnarray*}
with $\delta_{kk^\prime}$ the Kronecker delta.
Note that $\widehat{\mathrm{cov}}(\mathbf{\widehat{P}}^{*}(s,t))$ and $\widehat{\mathrm{cov}}(\Delta\mathbf{\widehat{\Lambda}}^{*}(t))$ are two $(K+1)^2 \times (K+1)^2$ covariance matrices.

These results may be used to construct the 95\% pointwise confidence intervals for the Aalen-Johansen estimator:
$$
\exp \left( \log \widehat{P}^{*}_{hk}(s,t) \pm 1.96 \dfrac{\sqrt{\widehat{\mathrm{var}}(\widehat{P}^{*}_{hk}(s,t))}}{\widehat{P}^{*}_{hk}(s,t))} \right).
$$

\subsection*{Appendix C: Example of R Code}

This example is based on the simulation study (see Section 4 in the article), with a non-homogeneous Markov multi-state model which included three states and three transitions.
The same covariate, called $X$ in the above code, impacted the longitudinal and survival processes, the longitudinal sub-model had a random intercept and a random slope, the log baseline intensities were approximated using B-splines, and the dependence between the two processes was explained through the true current level and the true current slope of the biomarker.

The \texttt{JMstateModel()} function and several detailed examples are available on \path{https:// github.com/LoicFerrer/JMstateModel/}.

\renewcommand{\baselinestretch}{1}
\begin{verbatim}
# Load the packages and the function to estimate joint multi-state models:
library(mstate)
library(JM)
source("JMstateModel.R")

# Import two databases which contain longitudinal and survival data:
load("data.RData")


###############################
#### Longitudinal sub-part ####
###############################

# Linear mixed model:
lmeFit <- lme(fixed = Y ~ (1 + times) * X,
              data = data_long,
              random = ~ (1 + times) | id,      
              method = "REML",
              control = list(opt = "optim"))


##############################
#### Multi-state sub-part ####
##############################

# Construct the 3*3 matrix of possible transitions:
tmat <- matrix(NA, 3, 3)
tmat[1, 2:3] <- 1:2
tmat[2, 3] <- 3
dimnames(tmat) <- list(from = c("State 0", "State 1", "State 2"),
                       to = c("State 0", "State 1", "State 2"))
tmat
# The transition '0 -> 1' is called '1','0 -> 2' is called '2' and
# '1 -> 2' is called '3'.

# Define the covariate(s) in the multi-state sub-part:
covs <- "X"

# The 'msprep()' function divides the survival database in order to have
# one line per transition at risk for each subject, with 'Tstart' the 
# entry time in the current state, and 'Tstop' the time of transition or
# censorship; 'status' denotes if the transition has been performed:
data_mstate <- msprep(time = c(NA, "time_of_state_1", "time_of_state_2"),
                      status = c(NA, "State 1", "State 2"),
                      data = data_surv,
                      trans = tmat,
                      keep = covs,
                      id = "id")

# 'expand.covs()' permits to define the set of covariates which impacts
# each transition:
data_mstate <- expand.covs(data_mstate, covs,
                           append = TRUE, longnames = FALSE)

# Multi-state model with proportional hazards:
coxFit <- coxph(Surv(Tstart, Tstop, status) ~
                  X.1 + X.2 + X.3 + strata(trans),
                data = data_mstate,
                method = "breslow",
                x = TRUE,
                model = TRUE)


####################################
#### Joint multi-state sub-part ####
####################################

# Define the derivative of the fixed and random parts in the mixed model,
# and indicate which covariates are kept, for the dependency on the slope
# of the marker:
dForm <- list(fixed = ~ 1 + X,
              indFixed = c(2, 4),
              random = ~ 1,
              indRandom = 2)

# Joint multi-state model with:
# 	- current level and current slope as dependence function,
# 	- adaptative cubic B-splines to approximate the log-baseline intensities,
# 	- 9 Gauss-Hermite quadrature points in the pseudo-adaptative numerical
# 	 		integration to approximate the integrals over random effects,
# 	- 3 Gauss-Kronrod quadrature points to approximate the integral over time.
jointFit <-
  JMstateModel(lmeObject = lmeFit,
               survObject = coxFit,
               timeVar = "times",
               parameterization = "both",
               method = "spline-PH-aGH",
               interFact = list(value = ~strata(trans) - 1,
                                slope = ~strata(trans) - 1,
                                data = data_mstate),
               derivForm = dForm,
               Mstate = TRUE,
               data.Mstate = data_mstate,
               ID.Mstate = "id",
               control = list(GHk = 9, lng.in.kn = 3))
summary(jointFit)

\end{verbatim}

\section*{Acknowledgements}

The authors thank Paul Sargos and Pierre Richaud from the Institut Bergoni\'e (Bordeaux, France) for their availability and their expertise in clinical interpretations.
Computer time for this study was provided by the computing facilities MCIA (M\'esocentre de Calcul Intensif Aquitain) of the Universit\'e de Bordeaux and of the Universit\'e de Pau et des Pays de l'Adour.
This work was supported by a joint grant from INSERM and R\'egion Aquitaine, and a grant from the Institut de Recherche en Sant\'e Publique [grant AAP12CanBio16].
The RTOG trial and J. Dignam's efforts were supported by Public Health Service grants U10 CA21661 and U10 CA180822 from the National Cancer Institute, NIH, U.S. Department of Health and Human Services.

\vspace*{-8pt}

\section*{Supplementary Materials}

The function \texttt{JMstateModel()}, which is an extension of the standard \texttt{jointModel()} function to the multi-state framework, is available with several examples on \path{https://github.com/ LoicFerrer/JMstateModel/}.\vspace*{-8pt}

\bibliographystyle{biom} 
\bibliography{JMstate}

\label{lastpage}

\end{document}